\documentclass[useAMS,usenatbib,onecolumn]{mn2e}
\usepackage{epsfig,amssymb,amsmath,cases,graphicx,enumitem,dsfont,mathrsfs,tablefootnote}

\newcommand{\be}{\begin{equation}}
\newcommand{\ee}{\end{equation}}

\newcommand{\mstar}{M_*}

\newcommand{\msun}{M_\odot}

\newcommand{\bsn}{\boldsymbol{\nabla}}
\newcommand{\bsb}{\boldsymbol{B}}
\newcommand{\bsF}{\boldsymbol{F}}

\title[Spin paramagnetism and NS deformations]{Spin paramagnetic deformation of a neutron star}
\author[A. G. Suvorov, A. Mastrano, and A. Melatos]{A. G. Suvorov\thanks{E-mail:
suvorova@student.unimelb.edu}, A. Mastrano\thanks{E-mail: alpham@unimelb.edu.au}, and A.
Melatos\thanks{E-mail: amelatos@unimelb.edu.au}\\School of Physics, University of Melbourne, Parkville VIC
3010, Australia}

\begin{document}

\date{Accepted ?. Received ?; in original form ?}

\pagerange{\pageref{firstpage}--\pageref{lastpage}} \pubyear{?}

\maketitle

\label{firstpage}

\begin{abstract}

\noindent{Quantum mechanical corrections to the hydromagnetic force balance equation, derived from the microscopic Schr\"{o}dinger-Pauli theory of quantum plasmas, modify the equilibrium structure and hence the mass quadrupole moment of a neutron star. It is shown here that the dominant effect --- spin paramagnetism --- is most significant in a magnetar, where one typically has $\mu_{B}|\bsb|\gtrsim k_B T_e$, where $\mu_{B}$ is {{the Bohr magneton}}, $\bsb$ is the magnetic field, and $T_e$ is the electron temperature. The spin paramagnetic deformation of a nonbarotropic magnetar with a linked poloidal-toroidal magnetic field is calculated to be up to ${{\sim 10}}$ times greater than the deformation caused solely by the Lorentz force. {{It depends on the degree of Pauli blocking by conduction electrons and the propensity to form magnetic domains, processes which are incompletely modelled at magnetar field strengths.}} The star becomes more oblate, as the toroidal field component strengthens. The result implies that existing classical predictions underestimate the {maximum} strength of the gravitational wave signal from rapidly spinning magnetars at birth. {Turning the argument around, future gravitational-wave upper limits of increasing sensitivity will place ever-stricter constraints on the physics of Pauli blocking and magnetic domain formation under magnetar conditions.}}

\end{abstract}


\section{Introduction}

Sufficiently strong internal magnetic fields deform neutron stars to the point where they may be detectable as gravitational wave sources by ground-based, long-baseline interferometers like the Laser Interferometer Gravitational Wave Observatory (LIGO) \citep{cutler02,mp05,hetal08,dss09,metal11,detal14}. The deformation is normally calculated within the theory of ideal magnetohydrodynamics (MHD) and is produced by the Lorentz force $\boldsymbol{J}\times \bsb$, where $\bsb$ is the magnetic field strength, and $\boldsymbol{J}=\mu_0^{-1}\bsn\times \bsb$ is the self-consistent current density \citep{cf53,g72,k89}. If the star is barotropic, such that pressure and density correspond one to one, the resulting MHD equilibria are solutions of the Grad-Shafranov equation. If the star is non-barotropic, e.g., due to entropy or lepton fraction gradients \citep{rg92,r01,r09}, a greater range of MHD equilibria can be constructed and analysed, such as the linked poloidal-toroidal `twisted torus' configuration found in numerical simulations \citep{bn06,b09,metal11,aetal13,detal14,msm14}. Furthermore, if neutron stars contain superfluid neutrons and superconducting protons \citep{migdal59,gk68,bpp69}, the Lorentz force changes its vector character to include terms like $\bsb\cdot\bsn(H_{c1}\hat{\boldsymbol{B}})$ \citep{ep77,r81}, where $H_{c1}\sim 10^{11}$ T is the type-II superconductor characteristic field strength, and the mass quadrupole moment scales as $BH_{c1}$ rather than $B^2$ \citep{aw08,gas11,l13,l14}.

MHD equilibria are important to calculate for several reasons. As in this paper, they are a key input into calculations of the star's {mass ellipticity} and hence its gravitational wave luminosity \citep{cutler02,setal05,dss09}. In addition, they are a starting point for assessing the long-term stability of the magnetic field and predicting energy releases during magnetar bursts. Field stability can be tested by numerical simulations \citep{bn06,b09,arv14} or analytically \citep{aetal13}. Magnetar burst observations, in particular the 1998 August 27 flare from SGR 1900$+$14, offer strong evidence for the existence of strong internal magnetic fields, and the bursts themselves can be interpreted as transitions between internal magnetic states \citep{i01,co11}. MHD equilibria also serve as a starting point for simulations of Hall drift \citep{mar14}. It is timely, therefore, to take into account as many realistic physical effects as possible when modelling neutron stars in hydromagnetic equilibrium.

Previous calculations of the magnetic deformation of a neutron star, including those referenced above, have neglected quantum mechanical corrections to the MHD equations of motion (with the important exception of superconductivity). In many applications, this is entirely justified. In most ordinary neutron stars, for example, the surface dipole magnetic field strength does not exceed the critical value $B_c = m_e^2 c^2/e\hbar=4.4\times 10^9$ T, where quantum electrodynamic processes are activated \citep{melrose}. Unless the internal field is much stronger than the observed dipole, spin-related effects are washed out by thermal fluctuations in these objects. Moreover many-body quantum correlations, which can lead to macroscopic effects in principle, are typically nullified in the hydrodynamic regime in practice (again, with the important exception of superconductivity). In magnetars, however, $B_c$ is comfortably exceeded, and the thermal energy per dynamical degree of freedom is less than the Larmor energy. Quantum kinetic effects, especially those involving spin, become important under such conditions and modify the `macroscopic' MHD equations \citep{degs1,bromar2}. The hydrodynamic description of quantum plasmas, including macroscopic spin effects (Haas, Manfredi, \& Feix 2000; Brodin \& Marklund 2007, hereafter BM07; Marklund \& Brodin 2007), enjoys a variety of applications to multi-stream instabilities \citep{aetal02}, linear and nonlinear ion-acoustic waves \citep{hetal03}, four-wave interactions and nonlinear Zakharov wave collapse \citep{getal05}, quantum dusty plasmas \citep{ss06}, laser-plasma experiments \citep{mtb06}, and fusion plasmas \citep{ckv86}.


In this paper, we calculate how one particular quantum effect --- spin paramagnetism --- modifies the MHD equilibrium structure of a magnetar. We focus on spin paramagnetism, because it is the leading quantum force correction relevant to magnetars within the framework of spin MHD (BM07). It sets the stage for future analyses of other quantum corrections, which lie outside the scope of this paper. In Sec. 2.1 and Appendix A, we review briefly the hydrodynamic description of quantum plasmas and its self-consistent derivation from the underlying, microscopic, quantum theory. In Sections 2.2 and 2.3, we estimate the order of magnitude of the spin paramagnetic corrections in a magnetar {{and some possible saturation mechanisms}}. In Sec. 3, we construct, as a worked example, a modified MHD equilibrium for a nonbarotropic star with a linked poloidal-toroidal magnetic field akin to those seen in simulations (Sec. 3.1) and calculate its {ellipticity} (Sec. 3.2). The results are compared with previous classical calculations in Sec. 4. The consequences for gravitational radiation are discussed briefly. {Throughout this work we employ SI units.}

\section{Quantum force density}

\subsection{Spin paramagnetism}

The multi-fluid equations of motion for a quantum plasma can be derived systematically from the $N$-particle Schr\"{o}dinger-Pauli equation including spin by following the prescription in BM07. The main steps in the derivation are summarised in Appendix A. In brief, for each species, one applies a Madelung transformation to the Schr\"{o}dinger-Pauli equation to write the amplitude of the $N$-particle wavefunction in terms of number density and the gradient of the phase in terms of bulk velocity. The number density and bulk velocity are averaged over the $N$-particle ensemble, with each term weighted by the corresponding single-particle probability. The $N$-particle wavefunction factorizes, if entanglement is negligible (i.e., if the decoherence time is short). The resulting equations of motion are given by equations (A6)--(A8) in Appendix A for each species. The terms on the right-hand sides are too complicated to write out in full; their forms are given explicitly by BM07. Physically, as well as the standard classical pressure and electromagnetic forces, there are macroscopic quantum mechanical forces arising from spin-spin self-correlations and correlations between spins and thermal velocity fluctuations. Related spin-spin and spin-thermal torques are exerted on the ensemble-averaged spin vector.

If the plasma is quasineutral, equations (A6)--(A8) for the protons and electrons can be combined into a single-fluid description like in classical MHD. Letting $\rho$, $p$, and ${\bf{v}}$ be the mass density, pressure, and bulk velocity {{of the single MHD fluid}} respectively, the momentum equation reads

\be \label{eq:forcebal}
\rho \left( \frac {\partial} {\partial t} + \boldsymbol{v} \cdot \bsn \right) \boldsymbol{v} = \boldsymbol{J} \times \boldsymbol{B} - \rho \bsn \Phi - \bsn P + \boldsymbol{F}_{Q}.\ee
{{In a neutron star,}} the forces in (\ref{eq:forcebal}) are felt {{specifically}} by the MHD proton-electron fluid, which typically amounts to a few per cent of the star's mass. {{In what follows, therefore, we approximate $\rho$ by the proton mass density $\rho_p$ when applying equation (\ref{eq:forcebal}).}} The first three terms on the right-hand side of (\ref{eq:forcebal}) are classical, representing the Lorentz force, the gravitational force (Newtonian potential $\Phi$), and pressure gradient respectively (the anisotropic pressure tensor is dropped for simplicity).


The last term on the right-hand side of equation (\ref{eq:forcebal}) is the quantum force density $\bsF_Q$. In general, it takes a complicated form [see equation (22) of BM07] involving spin-spin interactions and spin-thermal coupling. The latter effect enters implicitly through the spin transport equation (A11). However, in the MHD limit, where the Larmor radius is small compared to the length-scale of magnetic gradients, one can neglect the spin-spin, spin-thermal, and spin inertia terms to a good approximation (see Appendix A and BM07), and the quantum force density {{acting on the MHD proton-electron fluid (cf. Sec. 3)}} reduces to

\begin{equation} \label{eq:bromar44}
\bsF_{Q} = \frac{\rho_{p}} {m_p} \left[ \bsn \left( \frac {\hbar^2} {2 m_{p} \rho_{p}^{1/2}} \bsn^2 \rho_{p}^{1/2} \right) + G\left(| \bsb|, T_{e} \right) \mu_{B} \bsn | \bsb | \right],
\end{equation}
where $m$ denotes the particle mass, $\mu_{B}=e\hbar/(2m_e)$ is the Bohr magneton, $T$ is the temperature, $G(|\bsb|, T_e)$ is the Brillouin function, the subscripts $p$ and $e$ refer to the proton and electron components respectively, {{and  $\rho_{p}$ $(\gg \rho_e)$ is the proton mass density.}} {We assume that the particles are non-relativistic, like BM07, although this may not be true closer to the core of a neutron star. {The Brillouin function is a thermodynamic factor that equals the ratio $(n_{0+}-n_{0-})/(n_{0+}+n_{0-})$, where $n_{0+}$($n_{0-}$) are the number densities of particles in the higher (lower) spin states. The Brillouin function is given in full generality by equation (60) in \citet{zmb10} and contains the effects of Landau quantization, spin splitting, and Fermi-Dirac statistics. We consider the limits of this expression in Sections 2.2 and 2.3.}

The first term in equation (\ref{eq:bromar44}) is often referred to as the quantum pressure. It arises physically from the self-attraction between bosons. It dominates at small length-scales and is important in the study of superfluid vortex structure \citep{don91}. The second term represents a collective form of spin paramagnetism. In the regime $\mu_{B}|\bsb|\gtrsim k_B T_e$, the electron and proton spins tend to align with $\bsb$, with the electrons dominating the net polarization $(\mu_{B}\gg\mu_p)$. The spin-polarized proton-electron fluid therefore feels a paramagnetic force in the presence of a magnetic gradient.

{It is possible that inside neutron stars the quantum effects discussed here in the presence of strong magnetic fields also influence gravitational interactions. A relativistic and self-consistent analysis would then require a modification to the gravitational potential $\phi$ leading to additional expressions appearing in the quantum force $\boldsymbol{F}_{Q}$ \citep{dh80}.}


\subsection{Suppression of the paramagnetic force and magnetization}

{{The spin-MHD theory, developed by BM07, which is applicable when the length-scales of magnetic gradients exceed the Larmor radius (as in a neutron star), leads to a total magnetization given by equation (43) of BM07:

\be \label{eq:m}\boldsymbol{M}=\frac{\mu_B\rho_p}{m_p} G\left(|\bsb|, T_e\right) \hat\bsb.\ee

{In a neutron star, the electrons form a degenerate gas. Pauli suppression is expected to lower $|\boldsymbol{M}|$ by a factor of $T_e/T_{Fe}$, where $T_{Fe}$ is the Fermi temperature of the electrons $[T_{Fe}\sim 10^{12}(\rho_p/10^{15}\textrm{ kg}$ $\textrm{m}^{-3})^{2/3}\textrm{ K}]$, by analogy with the Pauli spin magnetization of the conduction electrons in a metal \citep{k66}, although the conduction band is modified strongly by the magnetic field at magnetar field strengths, with uncertain implications for $\boldsymbol{M}$. Thus, in the regime $T_e \ll T_{Fe}$ (relevant to this paper), the Brillouin function $G\left(|\bsb|, T_e\right)$, given by equation (60) of \citet{zmb10}, simplifies into}

\be G\left(|\bsb|, T_e\right) = \frac{3}{2}\left(\frac{\mu_B |\bsb|}{k_B T_{Fe}}\right),\ee
{which accounts for Pauli blocking. In contrast, in the high-temperature regime $T_e \gg T_{Fe}$, $G\left(|\bsb|, T_e\right)$ simplifies into}

\be G\left(|\bsb|, T_e\right) = \tanh\left(\frac{\mu_B |\bsb|}{k_B T_e}\right).\ee
{Additionally, for the simplified forms of $G(|\bsb|, T_{Fe})$ given by equations (4) and (5) to be valid, one must have $\mu_B |\bsb|<k_B T_{Fe}$, i.e., $|\bsb| \lesssim 1.5\times 10^{12}(\rho_p/10^{15}\textrm{ kg m}^{-3})^{2/3}$ T, which is readily satisfied in most neutron stars. Equations (3)--(5) imply a magnetization-to-magnetic-field ratio $\mu_0|\boldsymbol{M}|/|\bsb|\approx 7(\rho_p/10^{15} \textrm{ kg m}^{-3})$ $(T_{Fe}/10^{12}\textrm{ K})^{-1}$ for $T_e \ll T_{Fe}$. {Given $\boldsymbol{M}$, we can calculate the magnetic susceptibility $\chi=|\boldsymbol{M}|/|\boldsymbol{H}|$, where $\boldsymbol{H}$ is the magnetic induction, $\boldsymbol{H}=\bsb/\mu_0 - \boldsymbol{M}$. In the neutron star regime, where the conditions $T_e \ll T_{Fe}$ and $\mu_B |\bsb|< k_B T_{Fe}$ are satisfied, we find $\chi\approx 2\times 10^{-2}$.}}

In reality, the Brillouin model underpinning equation (3) may break down under neutron star conditions. The tendency towards alignment is reduced at high densities by the chemical potential, which is subtracted from $\mu_B |\bsb|$ in the Boltzmann probability leading to equation (3). {There are also subtle collective effects to consider. Equation (3) does not exhibit the expected oscillation of $\boldsymbol{M}$ versus $|\bsb|$, known as the de Haas-van Alphen oscillation, caused by changes in the number of occupied Landau levels \citep{bh82,fetal10,ns07,cetal15}. On the other hand, including the Heisenberg nearest-neighbour exchange interaction in the Brillouin theory tends to enhance spin alignment and encourage magnetic domains to form, a potentially strong effect which is nevertheless hard to quantify \citep{dzg13}. In this way, a more realistic expression for $\boldsymbol{M}$ includes a prefactor $K$ of the form \citep{k66}}

\begin{equation} \label{eq:mi6}
\boldsymbol{M}=K  \frac{\mu_B\rho_p}{m_p} G\left(|\bsb|, T_e\right) \hat\bsb,
\end{equation}
{where $K$ is a parameter which describes additional factors that are potentially absent from equation (60) of \citet{zmb10}}. Gravitational wave limits could be used to constrain the parameter $K$, since the deformation induced by the paramagnetic force will be directly proportional to the magnetisation \eqref{eq:mi6}.}

\begin{table*}
\begin{minipage}{\textwidth}
\centering

  \caption{{Selected} theoretical values of neutron star magnetic susceptibility $\chi$ {drawn from the literature.}}
  \begin{tabular}{lccc}
  \hline
    Model & $T$ & $|\boldsymbol{B}|$ & $|\chi|$\\
\hline
Relativistic mean-field{\footnote{\cite{dzg13}}} & Zero & $10^8\textrm{ T}<|\bsb|<10^{11}\textrm{ T}$ & $<20$\\
 & & \\
Colour-flavour-locked superconductivity\footnote{\cite{ns07}} & Low & $|\bsb|\lesssim 8.5\times 10^{15}$ T  & $\lesssim 1$\\
 &  & \\
Quantum chromodynamics with isospin chemical potentials\footnote{\cite{e14}} &Wide range & Wide range & $<0.1$\\
 & & \\
Relativistic mean-field theory with nonlinear meson interaction\footnote{\cite{retal14}} & Zero & $10^{11}\textrm{ T}<|\bsb|<10^{15}$ T & $<0.02$\\
&  & \\
Relativistic degenerate electron gas\footnote{\cite{s12}} & Wide range & $|\bsb|\lesssim 10^{13}$ T & $\sim 10^{-3}$\\

\hline
\end{tabular}
\label{table1}
\end{minipage}
\end{table*}


{We present some examples of theoretical values of $\chi$  \citep{ns07,s12,dzg13,e14,retal14} in Table 1 for comparison. As evident, there are disagreements regarding $\chi$ and the method of calculating it. We see also that our value for susceptibility, $\chi\sim 10^{-2}$, in the $\mu_B|\bsb|/k_B T_e\gtrsim 1$ regime is not very different from the values calculated using the methods summarized in Table 1.}

{All this suggests that the form of $\bsF_Q$ given by equation (44) of BM07 and equation (\ref{eq:bromar44}) in this paper is not the whole story (especially for the low-$|\bsb|$/high-$T$ regime), but it is hard to be confident about the form and order-of-magnitude of the corrections in the absence of experimental guidance, when the material and physical conditions are so exotic.} A self-consistent analysis of spin paramagnetism in a dense, highly-magnetized, multi-species fluid, as in a realistic neutron star, including collective effects and domain formation, lies beyond the scope of this paper. Our aim here is to point out that this force contributes to the stellar ellipticity and to estimate the maximum size of this contribution. The reader should bear in mind that the effect may be suppressed (or enhanced, in the case of domain formation) by the mechanisms in this paragraph and potentially others not referenced here. {Some mathematically consistent remedies which may lead to a different expression for the net magnetisation $\boldsymbol{M}$, and hence an alternate $\chi$ for the low-$|\bsb|$/high-$T$ regime, are explored in the Appendix.}

\subsection{Order-of-magnitude estimates}

Before constructing an MHD equilibrium explicitly as an example, we compare the characteristic magnitudes of the two quantum corrections in equation (\ref{eq:bromar44}) with the perturbing Lorentz force considered in previous classical analyses of neutron star deformations.

The ratio $\kappa_1$ of the quantum pressure to the Lorentz force is

\begin{eqnarray}
\kappa_1 &\sim& \frac{\mu_0 \hbar^2 \rho_{p}}{m_p m_e R^2 B^2}\\
&\sim& 9\times 10^{-33}  \left(\frac{\rho_{p}}{10^{15}\textrm{ kg m}^{-3}}\right)\left(\frac{R}{10^4\textrm{ m}}\right)^{-2}\left(\frac{|\bsb|}{10^{11}\textrm{ T}}\right)^{-2},
\end{eqnarray}
up to factors of order unity, where we approximate $\bsn$ with $R^{-1}$, where $R$ is the stellar radius. As expected, $\kappa_1$ is tiny. The quantum pressure is only important in vortex cores with diameter $\lesssim 10^{-11}$ m.

{The ratio $\kappa_2$ of the paramagnetic force to the Lorentz force is}

\begin{eqnarray}
\kappa_2 &\sim& \frac{3}{2}\frac{\mu_B^2 \mu_0 \rho_p}{k_B m_p T_{Fe}}\\
&\sim& 7 \left(\frac{\rho_p}{10^{15}\textrm{ kg m}^{-3}}\right)\left(\frac{T_{Fe}}{10^{12}\textrm{ K}}\right)^{-1},
\end{eqnarray}
{up to factors of order unity. The estimates in equations (9) and (10) is made for the regime $T_e \ll T_{Fe}$, where $G(|\bsb|, T_e) = (3/2)(\mu_B |\bsb|/k_B T_{Fe})$ \citep{zmb10}.}

Equation (10) demonstrates {{three}} important points. First, for typical magnetar fields, the spin paramagnetic correction to the Lorentz force $\boldsymbol{J}\times \bsb$ is large \citep{bromar2}. The ratio $|\bsF_Q|/|\boldsymbol{J}\times\bsb|$ corresponds to the ratio of the electron Larmor energy to the thermal energy per degree of freedom, {modified by Pauli blocking (Section 2.2)}. Second, the correction is appreciable even with respect to the `background' nonmagnetic forces. For example, the ratio of the spin paramagnetic force to the pressure gradient $\bsn P\sim \bsn(\rho k_B T_e/m_p)$ evaluates approximately to $\mu_{B} B/(k_B T_e)$, which can exceed unity in a magnetar. {Third, the correction nominally remains relevant in ordinary neutron stars ($B\lesssim 10^8$ T), {since $\kappa_2$ only depends on $\rho_p$ and $T_{Fe}$.} This counterintuitive result essentially arises because neutron star matter is so dense that even the weak residual alignment of spins in the regime $\mu_B |\bsb|\lesssim k_B T_e$ is enough to produce a significant magnetization per unit volume. In any event, neither spin paramagnetism nor the Lorentz force produce astrophysically interesting ellipticities, $\gtrsim 10^{-6}$ (that is to say, interesting from the gravitational wave viewpoint), for $|\bsb|\lesssim 10^7$ T.}

\section{Worked example: nonbarotropic star with a linked poloidal-toroidal magnetic field}

In this section, we calculate the {ellipticity} produced by the quantum force density $\bsF_Q$ for a linked poloidal-toroidal field of the form considered by many authors \citep{bn06,b09,metal11,detal14}. The single-fluid form of $\bsF_Q$ in equation (\ref{eq:bromar44}) acts on the electron-proton MHD fluid. The neutron condensate is also deformed by the spin paramagnetic force, but the form of $\bsF_Q$ is different and more complicated [see equation (22) of BM07]. Luckily, the effect on the neutrons is of order $(n_n/n_e)(\mu_n/\mu_{B})\sim 10^{-2}$ times the effect on the protons, where $n_n$ ($n_e$) and $\mu_n$ ($\mu_{B}$) are the number density and magnetic moment of the neutrons (electrons) respectively. It is therefore negligible in magnetar ellipticity calculations.

\subsection{Modified MHD equilibrium}

Consider an idealised, spherically symmetric, hydrostatic equilibrium satisfying $\bsn P+\rho\bsn\Phi=0$ such that, in terms of a normalised radial coordinate $r$, the density and pressure profiles are given by \citep{metal11,aetal13}

\begin{equation}
\rho(r) = \frac {15 M_{\star}} {8 \pi R^3} (1 - r^2),\ee
\be P(r) = \frac {15 G M_{\star}^2} {16 \pi R^4} \left(1 - \frac {5} {2} r^2 +2 r^4 - \frac {r^6} {2} \right),
\end{equation}
where $M_{\star}$ is the stellar mass and $R$ is the radius. We close the system with the Poisson equation $\bsn^2 \Phi = 4 \pi G \rho$. This background `parabolic' density profile is chosen for analytic simplicity, but \citet{metal11} showed a posteriori that the induced ellipticity is within 5 per cent of that obtained for a more realistic, polytropic density profile.

Introducing the magnetic field as the source of the perturbation, we write $\rho \mapsto \rho + \delta \rho$ along with $P \mapsto P + \delta P$. We take a poloidal and toroidal decomposition of an axisymmetric magnetic field \citep{c56},

\begin{equation} \label{eq:bfield}
\bsb(r,\theta) = B_{0} \left[ \eta_{p} \bsn \alpha \times \bsn \phi + \eta_{t} \beta(\alpha) \bsn \phi \right] ,
\end{equation}
where the flux function $\alpha(r,\theta)$ sets the poloidal field structure, and the dimensionless parameters $\eta_{p}$ and $\eta_t$ define the relative magnitudes of the poloidal and toroidal components respectively. The function $\beta$, which defines the toroidal field, must be a function of $\alpha$ to ensure that the azimuthal component of the Lorentz force vanishes, since there is no other azimuthal force to balance it in the axisymmetric MHD equilibrium. We further assume a dipole magnetic field, for which we may take \citep{metal11}

\begin{equation}
\alpha(r,\theta) = \frac {35} {8} \left(r^2 - \frac {6 r^4} {5} + \frac {3 r^6} {7}\right) \sin^2\theta,\label{eq:alpha}
\ee
\be \beta(\alpha) =
\begin{cases}
(\alpha - 1)^2&\textrm{for }\alpha\geqslant 1,\\
0&\textrm{for }\alpha < 1,
\end{cases}\label{eq:beta}
\ee
The field given by equations (\ref{eq:bfield})--(\ref{eq:beta}) is chosen to ensure that

\begin{enumerate}[leftmargin=*]
\item $\bsb$ is symmetric about the $z$-axis;
\item the poloidal component of $\bsb$ is continuous everywhere;
\item the toroidal component of $\bsb$ is confined to a toroidal region $(\alpha\geqslant 1)$ inside the star around the neutral curve;
\item $\boldsymbol{J}=\mu_0^{-1}\bsn\times\bsb$ is finite and continuous everywhere inside the star and vanishes at the surface ($r=1$).
\end{enumerate}

Keeping terms linear in the density and pressure while employing the Cowling approximation $(\delta \Phi = 0)$, the force balance equation \eqref{eq:forcebal} now reads

\be \label{eq:perturbed}
  \frac{1}{\mu_0}(\bsn\times\bsb)\times \bsb + \frac{\mu_{B}\rho_{p}}{m_p}  G \left(| \bsb |, T_e \right) \bsn | \bsb | = \bsn \delta P + \delta \rho \bsn \Phi,
\ee
where we neglect the quantum pressure term ($\kappa_1\ll\kappa_2$). We solve for $\delta\rho$ by taking the curl of both sides of equation (\ref{eq:perturbed}), matching the $\phi$-components, and then integrating with respect to $\theta$ \citep{metal11,mlm13}. In general, for any deforming axisymmetric net force $\boldsymbol{F}$ on the left-hand side of (\ref{eq:perturbed}), we have

\be \frac{\partial\delta\rho}{\partial\theta}=-\frac{r}{R}\frac{\mathrm{d}r}{\mathrm{d}\Phi}(\bsn \times \boldsymbol{F})_\phi.\ee

\begin{figure}
\centerline{\epsfxsize=15cm\epsfbox{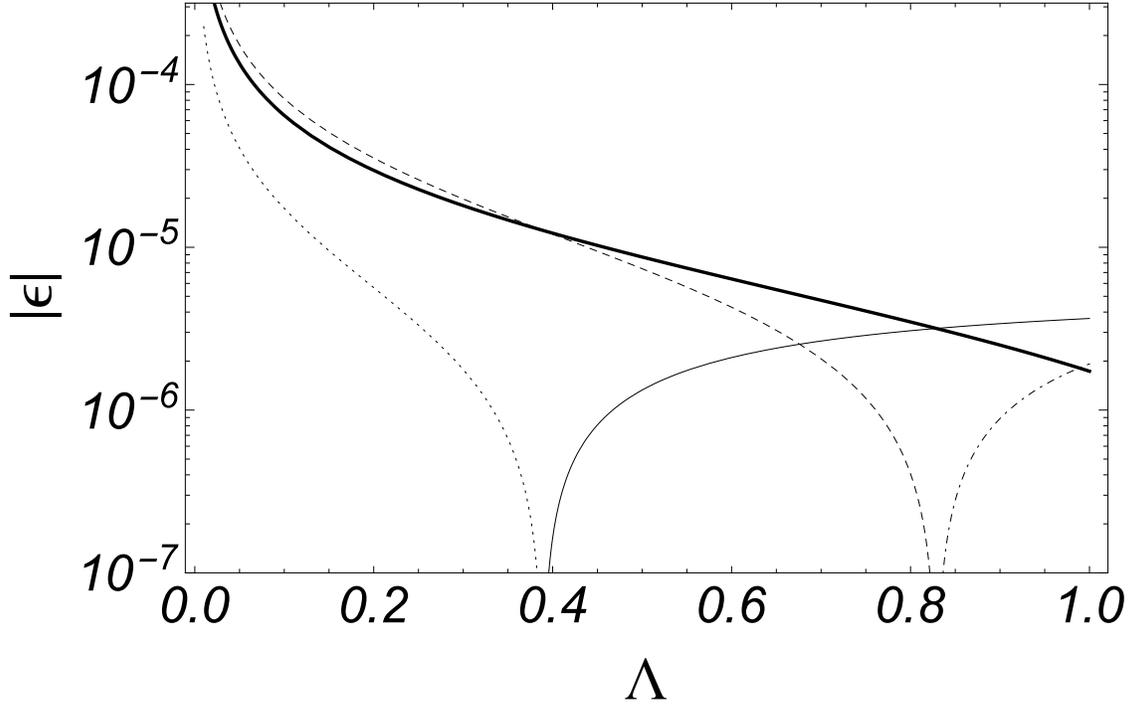}}
\caption{Logarithmic plot of the absolute value of mass ellipticity $\epsilon$ versus $\Lambda$, the ratio of poloidal to total magnetic field energy, for canonical magnetar parameters $B_0 = 5 \times 10^{10}$ T, $R = 10^{4}$ m, $M = 1.4 \msun$ and $T_e=10^{7}$ K. The thin solid curve corresponds to $|\epsilon_B|$, the deformation due solely to the Lorentz force exerted by a dipolar poloidal-toroidal field \citep{metal11}, for $\epsilon_B>0$. The thin dotted curve corresponds to $|\epsilon_B|$ for $\epsilon_B<0$. The thin dashed curve corresponds to $|\epsilon_Q|$, the deformation due solely to spin paramagnetism, for $\epsilon_Q>0$. The thin dashed-dotted curve corresponds to $|\epsilon_Q|$ for $\epsilon_Q<0$. $\epsilon_Q$ is directly proportional to $n_p/n_n$; we use $n_p/n_n = 10^{-3}$ here. The thick solid curve shows the total ellipticity $|\epsilon|=|\epsilon_B + \epsilon_Q|$. {Note that $\epsilon$ is always positive, but $\epsilon_B$ and $\epsilon_Q$ can be positive or negative.}}
\label{fig:boltzmann}
\end{figure}


\subsection{Ellipticity}

The stellar deformation is characterized by the mass ellipticity \citep{aetal08}

\be \epsilon = \frac{I_{zz}-I_{xx}}{I_0},\ee
where $I_0$ is the moment of inertia of the unperturbed spherical star, and the moment-of-inertia tensor is given by

\be I_{jk} = R^5 \int_V \textrm{d}^3x\,[\rho(r)+\delta\rho(r,\theta)](r^2\delta_{jk}-x_j x_k),\label{eq:mom}\ee
where the integral is taken over the volume of the star $(r\leqslant 1)$. We calculate $\epsilon$ by solving \eqref{eq:perturbed} for $\delta \rho$ and substituting into \eqref{eq:mom}. One picks up an integration constant when solving \eqref{eq:perturbed}, which is needed to ensure that $\delta\rho$ is continuous across $\alpha=1$. If the magnetic field contains higher-order multipoles, it is not always possible to find an integration constant that ensures continuity of $\delta\rho$, except in some special cases \citep{msm14}. Note that $I_0$ equals $n_n m_p R^5$ up to a multiplicative factor of order unity ($n_n\gg n_p$), whereas $I_{zz}-I_{xx}$ is proportional to $ n_p m_p R^5$ for the electron-proton MHD fluid described by equations (2) and (\ref{eq:perturbed}). Hence $\epsilon$ scales as $n_p/n_n$ overall. The neutrons are deformed by $\bsF_Q$ too, but this effect is small, as discussed in the first paragraph of Sec. 3.



In Fig. \ref{fig:boltzmann}, we plot $\epsilon_Q$ for $n_p/n_n=10^{-3}$ and $\epsilon_B$ versus $\Lambda=E_p/(E_p+E_t)$, where $\epsilon_Q$ is the ellipticity due solely to the spin paramagnetic effect, $\epsilon_B$ is the ellipticity due solely to a magnetic dipole field, and $E_{p}$ and $E_t$ are the {{total}} magnetic energy densities of the poloidal and toroidal components respectively. We plot $\epsilon_Q$ and $\epsilon_B$ separately to draw attention to their different behaviours; {{we also plot the the total ellipticity $\epsilon$, the sum of $\epsilon_Q$ and $\epsilon_B$, in the same figure}}. The Cowling approximation can change the value of $\epsilon_B$ by a factor of $\sim 2$ for the strongest magnetar fields \citep{y13}. We take the mass, radius, temperature, and surface equatorial magnetic field strength to be $1.4\msun$, $10^4$ m, $10^7$ K, and $5\times 10^{10}$ T respectively. {In the regime $T_e \ll T_{Fe}$, e.g. in a magnetar, $\epsilon_Q$ is independent of $T_e$.} The deformation $\epsilon_Q$ scales as $n_p/n_n$ if the deformation of the neutron fluid is neglected (see Sec. 3). For the canonical magnetar case shown in Fig. \ref{fig:boltzmann}, we derive

\be \epsilon_Q = -1.12\times 10^{-2} \left(\frac{n_p}{n_n}\right) \left(\frac{B_0}{5\times 10^{10}\textrm{ T}}\right)^2 \left(\frac{\mstar}{1.4\,\msun}\right)^{-1}\left(\frac{R}{10^4\textrm{ m}}\right)\left(1 - \frac{0.83}{\Lambda}\right).\label{eq:epsq}\ee

We see that $\bsF_Q$ {alone} tends to deform the star into a prolate shape for $\Lambda\gtrsim 0.83$ and into an oblate shape for $\Lambda\lesssim 0.83$. This is the opposite behaviour of the Lorentz force. We also see that, in general, the {maximum} spin paramagnetic deformation {(in the absence of Pauli blocking)} is greater than that caused by the Lorentz force ($|\epsilon_Q|>|\epsilon_B|$) for $\Lambda\lesssim 0.68$. Note however that equilibria with $\Lambda\lesssim 0.01$ are not expected to be stable \citep{b09,aetal13}.


Equation \eqref{eq:mom} describes the mass-density contribution to the moment of inertia arising from the $\rho c^2$ term in the $T^{00}$ component of the stress-energy tensor \citep{thorne1980}. In reality, there is also a direct electromagnetic contribution to $I_{jk}$ and hence $\epsilon$ arising from the $B^2/2\mu_0$ term in $T^{00}$; see equation (148) in Chapter 5 of \citet{degs1}. It can be shown, however, that this contribution only changes $\epsilon$ by about $\sim 2$ per cent \citep{msm14}. Note that $T^{\mu\nu}$ contains terms involving $\boldsymbol{S}$ in general but not in the $T^{00}$ component measured in the bulk frame.

\section{Discussion}

In this paper, we calculate the {ellipticity} of a strongly magnetized neutron star taking into account spin paramagnetism, the leading quantum mechanical correction to the MHD equations of motion. The maximum size of this correction exceeds the classical Lorentz force $(\boldsymbol{J}\times\bsb)$, when the condition $\mu_{B}|\bsb|>k_B T_e$ is satisfied, as routinely occurs in a typical magnetar. The correction arises physically because the electron spins (and, to a lesser extent, the proton spins) polarize the plasma, which then feels a force in a magnetic gradient. {Whether or not the correction reaches its maximum under realistic, astrophysical conditions depends on poorly understood physics like magnetic domain formation in a magnetar, which we do not attempt to model here.}

As a worked example, we calculate the {maximum} ellipticity $\epsilon_Q$ due to $\bsF_Q$ versus the poloidal-to-total magnetic energy ratio $\Lambda$ for a canonical magnetar. The behaviour of $\epsilon_Q$ as a function of $\Lambda$ is markedly different from the classical case. For example, the star becomes more oblate as the toroidal field strength increases (i.e., as $\Lambda$ decreases), unlike in the classical case (shown as the thin solid and thin dotted curves in Fig. \ref{fig:boltzmann}). For $n_p/n_n=10^{-3}$, the star becomes oblate {under $\bsF_Q$ alone for $\Lambda\gtrsim 0.83$.} For most values of $\Lambda$, $\bsF_Q$ has a stronger effect on $\epsilon$ than the Lorentz force. {We find typically $\epsilon_Q \approx -1.12\times 10^{-2} (n_p/n_n) (B_0/5\times 10^{10}\textrm{ T})^2$ in the magnetar regime. In the Brillouin approximation used in this paper, the magnetization is independent of $T_e$ {in the regime $T_e \ll T_{Fe}$.} Specifically, therefore, the surface temperature of magnetars should not alter the deformation substantially. In weaker field stars, with $\mu_B |\boldsymbol{B}| \lesssim k_B T_{e}$, it may play a larger role. {We find that we need $(\rho_p/10^{15}\textrm{ kg m}^3)(T_{Fe}/10^{12}\textrm{ K})^{-1}\gtrsim 0.14$ to obtain $|\bsF_Q|>|\boldsymbol{J}\times \bsb|$.} {We remind the reader that the foregoing values of $|\bsF_Q|$ are maxima; in reality $|\bsF_Q|$ depends on {other collective effects (some of which are discussed in Sec. 2.2),} which are not completely modelled at magnetar field strengths.}}

{{Not only does $\bsF_Q$ increase $|\epsilon|$ by about one order of magnitude relative to $\boldsymbol{J}\times\bsb$ (Fig. \ref{fig:boltzmann}), it also changes the shape of the star. The Lorentz force alone yields a prolate star for $\Lambda\lesssim 0.38$ and an oblate star for $\Lambda\gtrsim 0.38$. {Adding $\bsF_Q$ leads to an oblate star for all $\Lambda$.} \citet{cutler02} predicted that the wobble angle of a precessing prolate star with misaligned angular momentum and magnetic axes tends to grow, until these axes are orthogonal, which is the optimal state for gravitational wave emission. Thus, $\bsF_Q$, even as it increases $|\epsilon|$, may make detection of gravitational waves from magnetar-like sources more difficult.\footnote{The authors thank Bryn Haskell for bringing this possibility to our attention.}}}

Although it is not known with certainty that the protons in the interiors of neutron stars (and in particular magnetars) form a type II superconductor, there is circumstantial evidence for the thesis from X-ray measurements of the cooling rate of the central compact object in the supernova remnant Cassiopeia A \citep{yls99,hh09}. It is therefore worth comparing the spin paramagnetic force density with the standard superconducting terms like $\mu_0^{-1}\bsb\cdot\bsn(H_{c1}\hat{\bsb})$, which exceed $\boldsymbol{J}\times\bsb$ by a factor $H_{c1}/B$. From equation (95) of \citet{gas11}, {we find $|\bsF_Q|/|\mu_0^{-1}\bsb\cdot\bsn(H_{c1}\hat{\bsb})|\sim (3/2)\mu_0\mu_{B}^2\rho_{p}|\bsb|/(m_p k_B T_{Fe} H_{c1})\sim 7\times 10^2 (\rho_{p}/10^{17} \textrm{ kg m}^{-3})(T_{Fe}/10^{12}\textrm{ K})^{-1}(|\bsb|/H_{c1})$.} Hence the spin paramagnetic correction is comparable in magnitude to the modified Lorentz force in a superconductor for typical magnetar parameters with $H_{c1}\sim B$ {and $\rho_{p} \sim 10^{-3} \rho$}.

Because the signal-to-noise ratio $S/N$ of a gravitational wave source is directly proportional to $|\epsilon|$, we expect the spin paramagnetic force to enhance the $S/N$ of magnetars significantly (especially older, cooler ones), even for ranges of $\Lambda$ where $\epsilon_B$ is expected to be small. {For example, if $n_p/n_n=10^{-3}$, for $\Lambda=0.4$, we find $|\epsilon_B|=1.6\times 10^{-7}$ but $|\epsilon_Q|=1.2\times 10^{-5}$; for $\Lambda=0.3$, we find $|\epsilon_B|=1.8\times 10^{-6}$ but $|\epsilon_Q|=2\times 10^{-5}$.} The most likely magnetar gravitational wave source is a hot, newborn one, hypothesized to spin with an initial period $\sim 1$ ms \citep{td93,dss09}. As a quick example, consider a newborn magnetar in the Virgo cluster, rotating with initial spin period 0.97 ms and final spin period 10 s, with $B_0=5\times 10^{10}$ T \citep{dss09,metal11}. This magnetar has significant detectability ($S/N>10$) for $\Lambda\lesssim 10^{-2}$, which is the lower limit for stability \citep{b09,aetal13}. With $\bsF_Q$ in effect, however, one obtains $S/N>10$ with $\Lambda\lesssim 5\times 10^{-2}$ and $n_p/n_n=10^{-3}$. {Note that $\epsilon_Q$ is independent of temperature, as long as $T_e\ll T_{Fe}$ [see equation (A14) and \citet{zmb10}]; $T_{Fe}$ is indeed higher than the estimated birth temperature ($T_e\sim 10^{11}$ K) of neutron stars \citep{yp04,yetal04,dss09}.} {Gravitational-wave experiments of the kind above may play a role in constraining the uncertain Pauli suppression and domain formation physics described in Sec. 2.3 and Table 1 and may ultimately constitute the main application of our results. The limits implied by gravitational wave experiments with current detectors are above the $\chi$ values in Table 1, consistent with at least partial Pauli suppression, but stricter and more interesting limits will follow as gravitational wave detector sensitivities improve.}

The self-consistent hydrodynamic theory of a quantum plasma predicts the existence of several other quantum corrections arising from spin-spin and spin-thermal correlations, and from torques on the ensemble-averaged spin vector in the presence of a magnetic gradient. A cursory introduction to these effects is given in Appendix A, together with some key references, but their analysis lies outside the scope of this paper. {{There may also be nontrivial saturation physics which modifies the Brillouin model of the magnetization, as discussed in Sections 2.2 and 2.3.}} Our goal here is to alert the reader to the potential importance of quantum corrections and calculate a worked example for one leading effect, namely spin paramagnetism. {As discussed in Sections 2.2 and 2.3 and references therein, there are many theoretical and experimental uncertainties regarding the behaviour of matter in the high density, high magnetization regime. The simple Brillouin model used here simply gives the maximum magnetization one can expect. In future,} it will be interesting to extend the spin paramagnetic calculation to study different magnetic configuration, higher-order multipoles, and stability, as well as include some of the other quantum corrections discussed in BM07 {{and collective processes like the formation of magnetic domains}}.


\section*{Acknowledgments}

We thank Don Melrose for introducing us to key references in the quantum plasma literature, notably including BM07. {{We also thank Bryn Haskell, {the first reviewer} Taner Akg\"{u}n, {and the anonymous second reviewer} for their insightful comments, which have significantly improved the quality and clarity of this paper.}} This work was supported by an Australian Research Council Discovery Project Grant (DP110103347) and an Australian Postgraduate Award.

\appendix
\setcounter{secnumdepth}{1}
\section{Spin magnetohydrodynamics}

In this Appendix, we outline briefly the main steps involved in deriving the hydrodynamic quantum force density $\bsF_{Q}$ in equation \eqref{eq:forcebal} from the microscopic Schr\"{o}dinger-Pauli theory of a quantum plasma. The reader is referred to BM07 and references therein for a full treatment\footnote{{BM07 used SI units throughout their work but expressed the magnetic moment $\mu$ in terms of CGS units, resulting in the appearance of a factor of $c$, e.g. in equation (4) of BM07 and elsewhere.}}.

Consider an ensemble of $N$ nonrelativistic, spin-$\tfrac {1} {2}$ particles with mass $m$, magnetic moment $\mu$, and charge $q$, labelled by the index $\alpha$. Neglecting entanglement, as appropriate for a bulk fluid whose decoherence time is short, we can factorise the total system wavefunction $\Psi$ according to $\Psi = \prod^{N}_{\alpha = 1} \Psi_{(\alpha)}$, where the single-particle wavefunctions $\Psi_{(\alpha)}$ satisfy, {as per equation (9) of BM07},

\begin{equation} \label{eq:bromar9}
i \hbar \frac {\partial \Psi_{(\alpha)}} {\partial t} = \left[ - \frac {\hbar^2} {2 m} \left( \bsn - \frac {i q} {\hbar} \boldsymbol{A} \right)^{2} - \mu \bsb \cdot \boldsymbol{\sigma} + q \phi \right] \Psi_{(\alpha)}.
\end{equation}
In equation \eqref{eq:bromar9}, $\phi$ is the electric scalar potential, $\boldsymbol{A}$ is the magnetic vector potential (with $\boldsymbol{E} = - \bsn \phi - \partial \boldsymbol{A}/\partial t$ and $\bsb = \bsn \times \boldsymbol{A}$), and $\boldsymbol{\sigma} = (\sigma_{1}, \sigma_{2}, \sigma_{3})$ is a vector of Pauli matrices, with Cartesian components

\begin{equation} \label{eq:bromar3}
\sigma_{x} =  \left( \begin{array}{cc}
0 & 1 \\
1 & 0
\end{array} \right),
\hspace{8mm}
\sigma_{y} = \left( \begin{array}{cc}
0 & -i \\
i & 0
\end{array} \right),
\hspace{8mm}
\sigma_{z} = \left( \begin{array}{cc}
1 & 0 \\
0 & -1
\end{array} \right).
\end{equation}

To convert equation \eqref{eq:bromar9} into hydrodynamic form, we make a Madelung transformation
\begin{equation} \label{eq:bromar10}
\Psi_{(\alpha)} = n_{(\alpha)}^{1/2} \exp \left[ i S_{(\alpha)}/ \hbar \right] \varphi_{(\alpha)},
\end{equation}
where $n_{(\alpha)}$ is the number density of the $\alpha$-th particle, $S_{(\alpha)}$ is the phase,

\begin{equation} \label{eq:bromar13}
\boldsymbol{v}_{(\alpha)} = \frac {1} {m} \left[ \bsn S_{(\alpha)} - i \hbar \varphi^{\dag}_{(\alpha)} \bsn \varphi_{(\alpha)} \right] - \frac {q} {m} \boldsymbol{A}
\end{equation}
is the bulk velocity neglecting entrainment [cf. \cite{andbash,pca02,hask1}], $\varphi_{(\alpha)}$ is a two-component spinor, and

\begin{equation} \label{eq:bromar15}
\boldsymbol{S}_{(\alpha)} = \frac {\hbar} {2} \varphi^{\dag}_{(\alpha)} \boldsymbol{\sigma} \varphi_{(\alpha)}
\end{equation}
is the spin density vector. Upon substituting \eqref{eq:bromar10} into \eqref{eq:bromar9}, the Madelung transformation leads to seven coupled equations of motion representing conservation of particle number, conservation of momentum, and spin transport, given by equations (11), (12), and (17) respectively in BM07. In the hydrodynamic limits, these equations are averaged over the $N$-particle ensemble according to the prescription in section 2 of BM07, weighting terms in the average by the corresponding single-particle probability. The resulting equations of motion for the ensemble-averaged fields $n = \langle n_{(\alpha)} \rangle$, $\boldsymbol{v} = \langle \boldsymbol{v}_{(\alpha)} \rangle$, and $\boldsymbol{S} = \langle \boldsymbol{S}_{(\alpha)} \rangle$ can be written as

\begin{equation} \label{eq:bromar19}
\frac {\partial n} {\partial t} + \bsn \cdot ( n \boldsymbol{v} ) = 0,
\end{equation}
\begin{equation} \label{eq:bromar20}
m n \left( \frac {\partial} {\partial t} + \boldsymbol{v} \cdot \bsn \right) \boldsymbol{v} = q n ( \boldsymbol{E} + \boldsymbol{v} \times \boldsymbol{B} ) - \bsn \cdot \boldsymbol{\Pi} - \bsn P + \boldsymbol{F}_{Q},
\end{equation}
and

\begin{equation} \label{eq:bromar21}
n \left( \frac {\partial} {\partial t} + \boldsymbol{v} \cdot \bsn \right) \boldsymbol{S} = -\frac {2 \mu n} {\hbar} \boldsymbol{B} \times \boldsymbol{S} - \bsn \cdot \boldsymbol{K} + \boldsymbol{\Omega}_{s},
\end{equation}
In \eqref{eq:bromar19}--\eqref{eq:bromar21}, $\boldsymbol{\Pi}$ is the trace-free anisotropic pressure tensor, $P$ is the isotropic scalar pressure, $\boldsymbol{F}_{Q}$ is the quantum force density (discussed further below), $\boldsymbol{K}$ is the thermal-spin coupling tensor (ensemble averaged product of thermal velocity and spin perturbations), and $\boldsymbol{\Omega}_{s}$ is the nonlinear spin correction (ensemble averaged product of spin perturbations); see section 2 of BM07 for explicit definitions. If the fluid comprises multiple spin-$\tfrac {1} {2}$ species, the equations of motion take the form \eqref{eq:bromar19}--\eqref{eq:bromar21} for each species, with inter-species collision terms added to equation \eqref{eq:bromar20}.

In a quasineutral electron-proton plasma ($n_{e} \approx n_{p}$), equations \eqref{eq:bromar19}--\eqref{eq:bromar21} for the electron and proton fluids can be combined into a single-fluid description in the MHD limit, just like in a classical plasma but with spin transport added. The resulting single-fluid MHD equations of motion are

\begin{equation} \label{eq:conservation1}
\frac {\partial \rho} {\partial t} + \bsn \cdot ( \rho \boldsymbol{v} ) = 0,
\end{equation}
\begin{equation} \label{eq:forcebalance1}
\rho \left( \frac {\partial} {\partial t} + \boldsymbol{v} \cdot \bsn \right) \boldsymbol{v} = \boldsymbol{J} \times \boldsymbol{B} - \bsn \cdot \boldsymbol{\Pi} - \bsn P + \boldsymbol{F}_{Q} ,
\end{equation}
and

\begin{equation} \label{eq:bromar30}
\rho \left( \frac {\partial} {\partial t} + \boldsymbol{v} \cdot \bsn \right) \boldsymbol{S} = \frac {m_{e}} {e} \boldsymbol{J} \cdot \bsn \boldsymbol{S} - \frac {2 \mu \rho} {\hbar} \bsb \times \boldsymbol{S} - 2 m \bsn \cdot \boldsymbol{K} + 2 m \boldsymbol{\Omega}_{s}.
\end{equation}
In equations \eqref{eq:conservation1}--\eqref{eq:bromar30}, $\rho = m_{e} n_{e} + m_{p} n_{p} \approx m_{p} n_{p}$ denotes the total mass density,  $\boldsymbol{v} = \big( m_{e} n_{e} \boldsymbol{v}_{e} + m_{p} n_{p} \boldsymbol{v}_{p} \big)/\rho$ denotes the centre-of-mass velocity, $\boldsymbol{J} = -e n_{e} \boldsymbol{v}_{e} + e n_{p} \boldsymbol{v}_{p}$ denotes the total current density, subscripts $e$ and $p$ label the electron and proton species respectively, $\boldsymbol{\Pi} = \boldsymbol{\Pi}_{e} + \boldsymbol{\Pi}_{p} \approx \boldsymbol{\Pi}_{e}$ and $P = P_{e} + P_{p} \approx P_{e}$ are total pressure variables, $\boldsymbol{K}  \approx \boldsymbol{K}_{e}$ and $\boldsymbol{\Omega}_{s} \approx \boldsymbol{\Omega}_{s e}$ are dominated  by thermal-spin and spin-spin coupling within the lighter species, and $\boldsymbol{F}_{Q}$, the total quantum force density, is discussed further below. We neglect the anisiotropic pressure term $\boldsymbol{\bsn} \cdot \boldsymbol{\Pi}$ in this paper for simplicity.

Inside a neutron star, the length-scale of global magnetic gradients is much greater than the classical Larmor radius of the charges, so terms that are quadratic in $\boldsymbol{S}$ can be neglected in the expression for $\boldsymbol{F}_{Q}$ and also in equation \eqref{eq:bromar30}. Likewise, the spin-thermal coupling is small under these conditions. The spin inertia terms on the left-hand side of \eqref{eq:bromar30} are negligible, when the natural time-scale of the problem is much longer than the electron cyclotron frequency, e.g. for steady-state configurations. {The spin transport equation then implies $\boldsymbol{S} \times \bsb = \boldsymbol{0}$ and hence}

\begin{equation}
\boldsymbol{M} = \frac{ \rho_{p} \mu_{B}} {\hbar m_{p}} \boldsymbol{S}.
\end{equation}
{In the standard theory of paramagnetism, one has}
\begin{equation} \label{eq:bromar42}
\boldsymbol{S} = - \frac {\hbar} {2} G \left( | \bsb |, T_{e} \right) \hat{\bsb},
\end{equation}
{where $G\left( | \bsb |, T_{e} \right)$ is the Brillouin function}, $\mu_{B}$ is the electron magnetic moment (Bohr magneton), and $T_{e}$ is the electron temperature. {The Brillouin function $G\left( | \bsb |, T_{e} \right)$ is the ratio $(n_{0+}-n_{0-})/(n_{0+}+n_{0-})$ and is given by equation (60) of \citet{zmb10}. The form of $G\left( | \bsb |, T_{e} \right)$ simplifies, depending on the temperature regime \citep{zmb10}, into}

\begin{numcases}{G\left(|\bsb|, T_e\right)=}
\frac{3}{2}\left(\frac{\mu_B |\bsb|}{k_B T_{Fe}}\right)&$\textrm{for } T_e \ll T_{Fe}$,\label{ll}\\
\tanh\left(\frac{\mu_B |\bsb|}{k_B T_e}\right)&$\textrm{for } T_e \gg T_{Fe}$.\label{gg}
\end{numcases}
{The result in the regime $T_e\ll T_{Fe}$ (which is most relevant to this paper) accounts for Pauli blocking. For the simplified form $G=(3/2)(\mu_B |\bsb| /k_B T_{Fe})$ to be valid, one must also have $\mu_B |\bsb|<k_B T_{Fe}$, i.e., $|\bsb| \lesssim 1.5\times 10^{12}$ T. This latter condition is readily satisfied in most neutron stars.}



The general expression for $\bsF_{Q}$, which is the sum of equation (22) in BM07 for electrons and protons and contains complicated spin averages and spin-spin correlations, is too lengthy to write down here. However, it simplifies dramatically in the weak-gradient MHD limit above, where it takes the form

\begin{equation} \label{eq:bromar44b}
\bsF_{Q} = \rho \left[ \bsn \left( \frac {\hbar^2} {2 m_{p}^2 \rho^{1/2}} \bsn^2 \rho^{1/2} \right) +   G\left( \frac {\mu_{B} | \bsb|} {k_{B} T_{e}} \right) \frac {\mu_{B}} {m_{p}} \bsn | \bsb | \right].
\end{equation}
To a good approximation, $\bsF_{Q}$ for the electron-proton plasma is dominated by the quantum force on the lighter electrons, given by the two terms in \eqref{eq:bromar44b}. Neutrons also feel a quantum force density given by equation (22) of BM07. This effect is smaller (even though the neutrons are more abundant), as discussed in section 3. {We emphasise that $\rho$ in equation (A13) refers only to the charged component of the fluid, not the neutrons. It is therefore replaced by $\rho_p$ in equations (2) and (15) in the main body of the paper.}

\bsp \label{lastpage}

\end{document}